\documentstyle[twocolumn,epsf]{jpsj}

\def\runtitle{Chemical Brownian Motor}
\def\runauthor{Shin-ichi {\sc Sasa} and Tatsuo {\sc Shibata}}

\title{Brownian Motors driven by Particle Exchange}

\author{
Shin-ichi {\sc Sasa}\footnote{sasa@jiro.c.u-tokyo.ac.jp}
and Tatsuo {\sc Shibata}\footnote{shibata@complex.c.u-tokyo.ac.jp}
}

\inst{Department of Pure and Applied Sciences, University of Tokyo.\\
Komaba, Meguro-ku, Tokyo 153, Japan}

\recdate{\today; version 1.0}

\abst{We extend the Langevin dynamics  so that particles can be 
exchanged with  a particle reservoir. We show that  grand canonical 
ensembles are realized at equilibrium  and derive the relations of 
thermodynamics for processes between equilibrium states.
As an application of the proposed evolution rule, we  devise
a simple model of Brownian motors driven by particle exchange.}

\kword{Langevin Dynamics, Thermodynamics, Open Systems}

\begin{document}
\sloppy
\maketitle


\section{introduction}


Directed motion of a Brownian particle can be realized under
nonequilibrium conditions.  Recently, such Brownian motors have
been studied extensively, because they are believed to share
common features with biological molecular motors. \cite{As}
The oldest example of Brownian motors may be Feynmann's ratchet\cite{Fe} 
which works in contact with two heat baths, while the simplest example
may be made by considering time-depending potentials\cite{Ma,As2,Pr}. 
We then find another type of Brownian motors:
its  non-equilibrium nature is brought about by the existence of
two particle reservoirs with different chemical potentials.
This type of motors may be  most relevant to 
biological molecular motors. In particular, when we wish to study 
how chemical energy is converted to mechanical one,
we need to have a minimal model to be studied. 
Until now, some models, in which a chemical reaction is taken 
into account, have been proposed\cite{Ma2,As3,Zhou}. 
However, since the models of 
chemical reactions are assumed  without any energetic interpretations, 
they are not satisfactory to our motivation. We need a model
in which energetics associated with particle exchange 
can be discussed.


The time evolution of a position of a Brownian particle  $x$ is 
assumed to be described  by a Langevin equation
\begin{equation}
\gamma {d x \over dt}=-{\partial U \over \partial x}+\xi,
\label{lang}
\end{equation}
where $\gamma$ is a friction constant, $U(x)$ is a potential function,
and $\xi$ is thermal noise satisfying 
\begin{equation}
\langle \xi(t)\xi(t^\prime) \rangle =2 \gamma k_BT\delta(t-t^\prime).
\end{equation}
$k_B$ is a Boltzmann constant and $T$ denotes temperature. 
We focus on the regime where inertial effects are negligible. 
The energetic interpretation of the Langevin dynamics has been 
presented recently by Sekimoto.\cite{Ken} Defining the heat and the work for
nonequilibrium processes, he has found the first law of thermodynamics 
and  discussed the energetics for several examples including
Feynmann's ratchet.  Subsequently, Sekimoto and Sasa have shown
the second law of thermodynamics together with
a complementary relation between the irreversible heat 
and the time lapse.\cite{KS}
These results have revealed that the Langevin dynamics is a useful model
to discuss  energetic aspects of fluctuating systems. 
We thus wish to extend  the Langevin dynamics so that
we can discuss the energy transduction caused by particle exchange.  


The first step in the present study is to make a mathematical
model for equilibrium systems in contact with a particle reservoir.
In such systems, grand canonical ensembles should be realized. This
is the first necessary condition for the model we wish to have.  
Until now, two models, in which grand canonical ensembles are
realized, have been proposed. One is an extension of 
the Monte Carlo method,\cite{gcMC} 
and the other is designed as a deterministic system\cite{gcNH} on
the same methodology as Nos\'e invented.\cite{Nose,Hoover} 
 However, our concern is not restrict 
to equilibrium states, but includes non-equilibrium processes. 
We conjecture that the thermodynamic laws cannot be derived 
in previously proposed methods. In this paper, we present a model which 
satisfies two necessary conditions: the realization of grand canonical 
ensembles at equilibrium and the obedience of the thermodynamic laws
for processes between equilibrium states.


This paper is organized as follows.
In \S 2, we  study a lattice model and find an evolution rule 
which realizes grand canonical ensembles. We expect that a continuum 
limit of the lattice model has a certain relation with the Langevin 
dynamics in contact with a particle reservoir. In \S 3, relying 
on this implicit correspondence between them, we translate the evolution rule 
on the lattice model to  one appropriate to the Langevin dynamics.
The validity of our rule is confirmed by numerical simulations.
In \S 4, we derive the thermodynamic relations for processes
between equilibrium states. In \S 5, we devise a model 
of Brownian motors driven  by particle exchange between
particle reservoirs. The final section is devoted to discussion. 


Before closing this section, we mention a configuration of 
the system we study.  In order to avoid unnecessary complicatedness,
we analyze one dimensional systems defined in the region $x >0$. 
The boundary with a particle reservoir is assumed to be 
located at $x=0$. Further, interactions among particles are ignored.
That is, a potential force acting on each particle is 
determined by the particle position. In addition, the potential gradient
is assumed to vanish at the boundary with the particle reservoir, 
because we wish to neglect dynamical variables in particle reservoirs.
We also assume that the system contacts with a single heat bath of the
temperature $T$.

\section{lattice model}

The lattice we study consists of $M+1$ cites labeled by 
integers from 0 to $M$.  
A particle on the $i$-th cite can move to the adjacent cites,
the $i+1$-th and the $i-1$-th cite, with  the transition 
probability (per unit time) $w_{i,i+1}$ and $w_{i,i-1}$, respectively.
In the Langevin equation eq.(\ref{lang}), 
the transition probability from $x$ to $x+\Delta x$ during 
a time interval $\Delta t$ is proportional to\cite{Risken} 
\begin{equation}
\exp(- \gamma \beta{(\Delta x)^2 \over 4 (\Delta t)}
-\beta{\Delta U \over 2}),
\label{trans:lang}
\end{equation}
where $\beta=(k_BT)^{-1}$. 
Since  the diffusion expressed in the first term 
corresponds to a random walk procedure in the lattice model, 
the simplest form of $w_{ij}$ which has a relation with eq.(\ref{trans:lang})
may be given by 
\begin{equation}
w_{ij}=d \exp ( -{\beta \over 2} (U_j-U_i)) 
\label{trans:lattice}
\end{equation}
for $|i-j|=1$, where $d$ will be related to a diffusion constant.
The 0-th cite corresponds to a particle reservoir where 
the  density is kept at constant. We thus assume that $n_*$ particles 
always occupy the 0-th cite irrespective of the transition 
from/to the 1-st cite, where  $n_*$ will be related to the particle 
density in the reservoir.  The other cites compose the system in 
question.

Let us analyze the probability distribution $P(\{n_i \}, t)$, 
where $n_i$ denotes the particle number on the $i$-th cite. 
The evolution equation of $P$ is written as\cite{Kampen}
\begin{equation}
{\partial \over \partial t} P(\{n_i \}, t)=
\sum_{0 \le i,j \le M} w_{ij} 
(\hat a_{i}^\dagger \hat a_{j}-1) n_i P(\{n_i \}, t),
\label{evolv}
\end{equation}
where $n_0=n_*$, $\hat a_i^\dagger$ and $\hat a_i$ are creation and 
annihilation operators of a particle on the $i$-cite,
respectively. That is, they are defined as
\begin{eqnarray}
\hat a_i^\dagger 
f(n_1,n_2,\cdots,n_M)&=&f(\cdots, n_i+1, \cdots), \\
\hat a_i 
f(n_1,n_2,\cdots,n_M)&=&f(\cdots, n_i-1, \cdots),
\end{eqnarray}
and $\hat a_0^\dagger=\hat a_0=1$.
The stationary solution $P_s(\{n_i\})$ can be obtained in the
factorization form
\begin{equation}
P_s(\{n_i \})=\Pi_{i=1}^M 
{1 \over n_i !} \exp(-\bar n_i ) \bar n_i^{n_i}.
\label{statio}
\end{equation}
Actually, the substitution of this expression into eq.(\ref{evolv})
gives the detailed balance condition
\begin{equation}
w_{i,i+1} \bar n_i=w_{i+1,i} \bar n_{i+1},
\label{balance}
\end{equation}
where $ 0 \le i \le M-1 $ and $\bar n_0=n_*$.
Solving eq.(\ref{balance}),  we obtain
\begin{equation}
\bar n_i=n_* \exp(-\beta (U_i-U_0)).
\end{equation}
We then define the chemical potential $\mu$ as
\begin{equation}
\mu=U_{0}+k_BT \log n_*,
\label{cpot}
\end{equation}
by which eq.(\ref{statio}) is rewritten in the form
\begin{equation}
P_s(\{n_i \})={ 1\over \Xi}
{ 1\over \Pi_{i=1}^M  n_i !} 
\exp( \beta( \sum_{i=1}^M (\mu-U_i)n_i)).
\label{statio:2}
\end{equation}
Here, $\Xi$ is a normalization constant called the grand partition
function, and calculated as 
\begin{equation}
\Xi=\exp(\sum_{i=1}^M \exp(\beta(\mu-U_i))).
\label{gpart}
\end{equation}
The expression of eq.(\ref{statio:2}) shows that the particle
distribution is given by a grand canonical ensemble with 
the inverse temperature $\beta$ and the chemical potential $\mu$.
We can discuss  statistical properties based on eq.(\ref{statio:2}),
for instance,  the total number of particles in the system 
turns out to be poissonian with the average $\bar N$, where 
\begin{equation}
\bar N=\sum_{i=1}^M \bar n_i= \sum_{i=1}^M \exp(-\beta (U_i-\mu)).
\label{aveN}
\end{equation}

We next derive  the rate equation for the particle number on 
the $i$-cite, $\tilde n_i(t)$, which is defined as
\begin{equation}
\tilde n_i(t)=\sum_{\{n_i \}}  n_i P(\{n_i\}, t).
\label{aven}
\end{equation}
Multiplying eq.(\ref{evolv}) with $n_i$ and summing
$\{n_i \}$, we obtain 
\begin{eqnarray}
{d \tilde n_i \over d t}&=&
d \exp(-{\beta\over2}(U_{i}-U_{i+1}) )\tilde n_{i+1}
\nonumber \\
&+&d \exp(-{\beta\over2}(U_{i}-U_{i-1}) )\tilde n_{i-1}
\nonumber \\
&-&d[\exp(-{\beta\over2}(U_{i+1}-U_{i}))
\nonumber \\
&+&\exp(-{\beta\over2}(U_{i-1}-U_{i})) ]\tilde n_i
\label{rate:bulk}
\end{eqnarray}
for $1 \le i \le M-1$, and 
\begin{eqnarray}
{d \tilde n_M \over d t}&=&
d \exp(-{\beta\over2}(U_{M}-U_{M-1}) )\tilde n_{M-1}
\nonumber \\
&-&d\exp(-{\beta\over2}(U_{M-1}-U_{M})) \tilde n_M.
\label{rate:M}
\end{eqnarray}
The continuum limit of the expressions of eqs.(\ref{rate:bulk}) 
and (\ref{rate:M}) will be useful in the argument below. 
We introduce a lattice spacing $\delta x$ and assume the 
appropriate scaling of parameters and variables for $\delta x$:
$\rho(i\delta x)=\tilde n_i/\delta x $, 
$U(i \delta x)=U_i$, $x=i \delta x$, $L= M \delta x$, 
$\rho_*=n_* /\delta x$, and 
$d={k_BT/\gamma} /(\delta x)^{2}$. 
Then, the  limit $\delta x \rightarrow 0$  provides us 
\begin{equation}
{\partial \rho \over \partial t}=
{1 \over \gamma}{\partial \over \partial x} 
({\partial U \over \partial x}+k_BT{\partial \over \partial x})\rho,
\label{FP}
\end{equation}
and the boundary conditions
\begin{equation}
\rho(0,t)=\rho_*
\end{equation}
at $x=0_{}$ and the density flux vanish at $x=L$. 
Equation (\ref{FP}) has the same form as 
the Fokker-Planck equation. However, this is not an evolution 
equation for the probability distribution, but the rate equation 
for the particle number distribution. Its actual time evolution
fluctuates around a solution to the rate equation.  

The stationary solution of eq.(\ref{FP}) under the boundary
conditions is derived as 
\begin{equation}
\rho_s(x)=\rho_*\exp(-\beta (U(x)-U(0))),
\end{equation}
and  the chemical potential becomes
\begin{equation}
\mu=U(0)+k_B T \log (\rho_* \delta x).
\label{cpot:con}
\end{equation}
Here, $U(0)$ can be chosen arbitrary because
there is no absolute zero in a classical world.
Also, $\delta x$ is an unobservable parameter which 
may be identified to De Broglie length for thermal motion.
In the argument below, $U(0)$ and $\delta x$ will be 
assumed to be arbitrary constants.

\section{Langevin Dynamics}

In this section, we describe motion of Brownian particles 
in the  physical space $x > 0$  which contacts with a particle 
reservoir at $x = 0$.  The evolution equation for a particle position
is given by eq.(\ref{lang}). We solve this equation numerically 
by employing  a discretization scheme with a time step $\delta t$. 
The question here is to find a rule related to  absorbing and 
emission from/to the particle reservoir. Since we already know
the proper rule for the lattice model, we translate this rule
to suitable one for the Langevin dynamics. First, 
when a particle enters into the region $x \le 0$, this particle 
should be interpreted to be absorbed in the particle reservoir.
The emission rule is a little bit tactical. 
Recall that $n_*$  particles are always located 
at the $0$-th cite in the lattice model.
Since we wish to make a similar configuration, 
we assume the following rule: At each time step, a virtual particle 
is put randomly in the region $-1/\rho_* \le x \le 0$  and is moved 
by performing the $\delta t$ integration of eq.({\ref{lang}). 
Then, if the particle enters into the region $x >0$, this particle 
should be interpreted to be emitted from the reservoir.

We do not have a mathematical proof that grand canonical ensembles 
are realized by the rule given here.
Nevertheless, we can check its validity by numerical simulations.
As a simple example, we assume that an harmonic potential
$U= a x^2/2$ bounds particles around $x=0$. 
If our  model realizes the grand canonical ensemble 
with the chemical potential given by eq.(\ref{cpot:con}),
a total number of particles $N$ should obey the Poisson distribution:
\begin{equation}
P(N)={1\over N!} \exp(-\bar N) \bar N^N,
\label{gc}
\end{equation}
where $\bar N$ is calculated as
\begin{equation}
\bar N=\rho_* \sqrt{\pi k_B T \over 2a}.
\end{equation}
We performed numerical simulations with parameter values
$(a,\gamma,\beta,\rho_*,\delta t)=(1,1,1,1,0.01)$ and made
$P_{num}(N)$, a distribution function of the particle number, 
by using $10^4$ samples every one unit time. We then found
that $ {\rm sup}_N|P(N)-P_{num}(N)| < 5 \times 10^{-3}$.
We also measured how the averaged particle number 
$\bar N_{num}$ depends on the number of samples, $K$.
Figure 1 shows that $|\bar N_{num}-\bar N|$ decreases as $1/\sqrt{K}$,
but it has a plateau after $K > 5 \times 10^5$. 
It comes from the finiteness of $\delta t$. 
Actually, when $\delta t=0.001$, such a plateau was not observed
for $K < 10^6$. We believe that $\bar N_{num}$ approaches
to $\bar N$ in the limit $\delta t \rightarrow 0$ and 
$K \rightarrow \infty$. 
Therefore, we conclude that our evolution rule realizes the
grand canonical ensemble with the chemical potential eq.(\ref{cpot:con}).


\begin{figure}
\epsfysize=.5\textwidth
\centerline{\epsffile{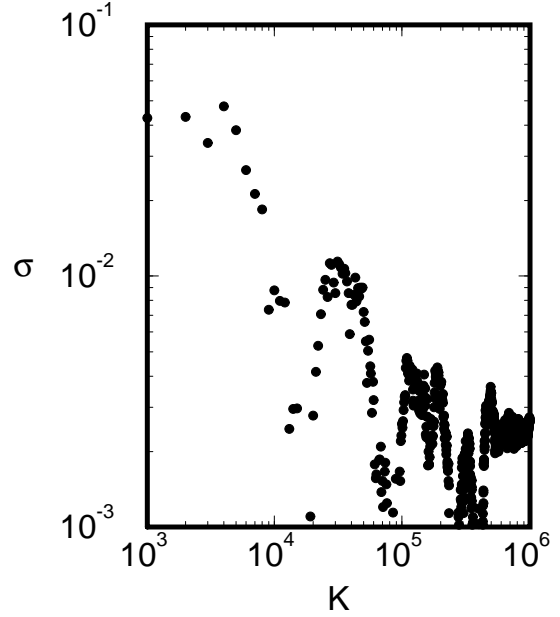}}
\caption{
$\sigma=|\bar N_{num}-N|$ versus the number of samples, $K$.
}
\label{fig1}
\end{figure}

We remark here how the indices of
particle positions are assigned in the theoretical analysis developed below.
Suppose that there are $N(0)$ 
particles in the system at $t=0$. Then, a position of the
$i$-th particle is denoted by $x_i(0) $, $(1 \le i \le N(0))$, 
where the ordering is assumed by some rule. When a particle
enters to the system first, the particle position is denoted by  
$x_{N(0)+1}$. Similarly, the unused minimum index is assigned 
to a position of the new particle. Further, when the $i$-th
particle exists in the particle reservoir, $x_i=0$ is assumed
so as to keep the continuity of $x_i(t)$. 
This convention will be useful, because we do not need to take care
whether a particle is in the system or in the reservoir. Instead,
we analyze an infinite number of particles. 

\section{thermodynamics}

In this section, we discuss energetics when an external
agent changes a parameter $\alpha$ of the potential $U$
during a time interval $[0,\tau]$. 
Since the external agent should not influence the particle
reservoir, we assume that $U(0)$ is kept at constant.
The change of the total potential energy 
in the course of the time evolution is expressed by
\begin{equation}
\Delta U_{tot}= \sum_{i=1}^{\infty}\int_0^\tau dt
\left[
 {\partial U(x_i,\alpha) \over \partial x_i}{d x_i \over dt}
+{\partial U(x_i,\alpha) \over \partial \alpha}{d \alpha \over dt}\right],
\label{utot}
\end{equation}
where the multiplication has been defined in the Stratonovich sense.
By using the evolution equation, the first term is rewritten as
\begin{equation}
Q_{tot}\equiv  \sum_{i=1}^{\infty}  \int_0^\tau dt
(-\gamma {d x_i\over dt}+\xi)\cdot {d x_i \over dt}
\label{heat}
\end{equation}
which can be identified with the energy transfer from 
the heat bath.\cite{Ken} The second term in eq.(\ref{utot}),
denoted by $W$, is interpreted
as the work done by the external agent. 
In this way, for each solution 
to the Langevin dynamics, we have an energy conservation law
\begin{equation}
\Delta U_{tot}=Q_{tot}+R.
\label{1st}
\end{equation}

We next discuss the energy transfer between the system and
the particle reservoir by assuming
\begin{eqnarray}
\Delta U_{tot} &=& \Delta U+\Delta U_R, 
\label{decom:U}\\
       Q_{tot} &=& Q+Q_R,
\label{decom:Q}
\end{eqnarray}
where the subscript $*_R$  denotes the contributions in the reservoir.
We do not know the way how to decompose $Q_{tot}$ based on the Langevin
dynamics. Instead, we employ the thermodynamic consideration as follows.
\begin{eqnarray}
\Delta U &=& Q+R+\mu \Delta N,
\label{1st:sys} \\
\Delta U_R &=& Q_R+\mu \Delta N_R,
\label{1st:res}
\end{eqnarray}
where note that the work cannot be extracted from the particle reservoir
because of the equality $\partial U/\partial \alpha=0$ at $x=0$.
Then, from  the equality
\begin{equation}
\Delta U_R=U(0)\Delta N_R=-U(0) \Delta N,
\label{delu:res}
\end{equation}
we obtain 
\begin{equation} 
Q_{R}=k_B T \log(\rho_*\delta x) \Delta N.
\label{qres}
\end{equation}
This implies that heat associated with particle exchange
flows to the particle reservoir. Therefore, the heat transferred
to the system from the other all region should be thought as $Q$, 
not $Q_{tot}$. 
We will find that such distinction becomes important when
we discuss a relation between quasi-static heat and the
thermodynamic entropy. (See eqs.(\ref{heatform}) and (\ref{kitahara}).)


Now, we discuss the second law of thermodynamics.
In the present case, this is represented by a minimum work principle:
the average of $W$ over the possible realizations of paths has a 
minimum value determined by a thermodynamic potential. 
In order to prove it, we employ the lattice model again,
in which the  average of $W$ is given by
\begin{equation}
\langle W \rangle =\int_0^\tau dt { d\alpha \over dt}
\sum_{\{n_i \}} \sum_{i=1}^M n_i { \partial U_i \over \partial \alpha}
P(\{n_i\}, t).
\end{equation}
Summing $\{ n_i \}$ first and recalling the definition of 
$\tilde n_i(t)$ given by eq.(\ref{aven}), we obtain 
\begin{equation}
\langle W \rangle =\int_0^\tau dt {d \alpha \over dt} 
\sum_{i=1}^M \tilde n_i(t){ \partial U_i \over \partial \alpha}.
\end{equation}
The continuum limit of this expression becomes
\begin{equation}
\langle W \rangle =\int_0^\tau dt {d\alpha \over dt} 
\int_0^\infty dx  \rho(x,t) {\partial U(x,\alpha) \over \partial \alpha},
\end{equation}
which takes the similar form as the averaged work 
for the case of  one particle Langevin dynamics.
Also, as mentioned above, $\rho$ satisfies
the Fokker-Planck equation. We thus can develop a similar
argument to the previous related study.\cite{KS} 
Introducing a stretched time  $s=t/\tau$ and the scaled variable
$\tilde \alpha(t/\tau)=\alpha(t)$, we expand $\rho$ and $\langle W
\rangle $
in such a way that 
\begin{eqnarray}
\rho &=& \rho_0+{1\over \tau} \rho_1+\cdots,\\
\langle W \rangle &=& \langle W \rangle_0+
{1\over \tau}\langle W \rangle_1+\cdots,
\end{eqnarray}
where we have assumed that $1/ \tau$ is a small parameter. 
We then solve eq.(\ref{FP}) perturbatically. At the lowest order,
we obtain 
\begin{equation}
\rho_0(x,s)=\rho_* \exp( -\beta (U(x,\tilde \alpha(s))-U(0))),
\end{equation}
and this yields 
\begin{eqnarray}
\langle W \rangle_0 &=& 
\int_0^1 ds {d \tilde \alpha (s) \over ds} 
\int_0^\infty dx \rho_0(x,s) 
{\partial U(x,\tilde \alpha) \over \partial \tilde \alpha}, \\
  &=& -k_B T [\bar N(\tilde\alpha(1))-\bar N(\tilde \alpha(0))].
\end{eqnarray}
Here, from eqs.(\ref{gpart}) and (\ref{aveN}), 
the grand potential defined by $\Omega=-k_B T \log \Xi $ 
is calculated as $-k_BT \bar N$. As the result, the quasi-static work 
turns out to be equivalent to the increment of the grand potential, that is,
\begin{equation}
\langle W \rangle_0= \Delta \Omega.
\label{grand}
\end{equation}
Therefore, in the quasi-static limit, eq.(\ref{1st:sys}) becomes
\begin{equation}
\Delta \bar U= \langle Q \rangle_0 +\Delta \Omega+\mu \Delta \bar N,
\label{1st:2}
\end{equation}
where $\bar U$ and $\bar N$ are the energy and the particle number 
averaged over the equilibrium ensemble for a given value of 
$\tilde \alpha$. 
Then, when we define the entropy through a thermodynamic relation
of the grand potential
\begin{equation}
\Omega  = \bar U-T S -\mu \bar N,
\label{1st:3}
\end{equation}
we recover a standard relation:
\begin{equation}
\langle Q \rangle_0=T \Delta S.
\label{heatform}
\end{equation}
{}Substituting eqs.(\ref{qres}) and (\ref{heatform}) to eq.(\ref{decom:Q}),
we also obtain 
\begin{equation}
\langle Q_{tot} \rangle_0
=T \Delta S+k_B T \log (\rho_* \delta x) \Delta \bar N.
\label{kitahara}
\end{equation}
Kitahara has proposed a similar expression in the context of
the work efficiency for  heat engines in open systems.\cite{kitahara}


At the  next order in the perturbative expansion, 
we obtain
\begin{equation}
\hat L \rho_1=-{\partial \rho_0 \over \partial s},
\label{linear}
\end{equation}
where the operator $\hat L$ is given by
\begin{equation}
\hat L=-{1\over \gamma} {\partial \over\partial x}
\left[ {\partial U \over \partial x}+
k_B T {\partial \over \partial x} \right].
\end{equation}
In order to solve eq.(\ref{linear}) under the boundary
condition $\rho_1(0)=0$, we introduce a Green
function defined by
\begin{equation}
\hat L( \rho_0(x) g(x,y))=\delta(x-y),
\label{green}
\end{equation}
where $g(x,y)=g(y,x)$ and $g(0,y)=0$. Such a Green function 
exists  owing to the Hermiteness of the operator
$\hat L(\rho_0  \cdot)$ defined under the given boundary condition.
Also, since we focus on the case that a stable stationary distribution
is realized,  all eigenvalues of $\hat L$ are expected to be non-negative, 
and this leads that $g(x,y)$ is a semi-positive definite in the sense that 
\begin{equation}
\int_0^\infty dx \int_0^\infty dyg(x,y)\phi_1(x)\phi_2(y) \ge 0,
\label{positive}
\end{equation}
where $\phi_1$ and $\phi_2$ are arbitrary functions in a certain 
functional space. 
Using the Green function $g(x,y)$, we express $\rho_1(x)$ as
\begin{equation}
\rho_1(x)=[-\int_0^\infty dy  g(x,y)
{\partial \rho_0 \over \partial s} ] \rho_0(x).
\end{equation}
This yields the first order contribution  to the non-static work:
\begin{eqnarray}
\langle W \rangle_1 & =& 
\int_0^1 ds 
\left( {d \tilde \alpha(s) \over ds} \right)^2 
\int_0^\infty dx  \int_0^\infty dy  \nonumber\\
&& {\partial U(x,\tilde \alpha) \over \partial \tilde \alpha}
g(x,y)
 {\partial U(y,\tilde \alpha) \over \partial \tilde \alpha}
\rho_0(x)\rho_0(y).
\end{eqnarray}
We then find that $\langle W \rangle_1 $ is non-negative
due to eq.(\ref{positive}). Therefore, the minimum work principle
\begin{equation}
\langle W \rangle \ge \langle W \rangle_0=\Delta \Omega
\end{equation}
holds within the validity of the  perturbation theory.
In the similar way as developed in the previous study,\cite{KS} 
we can also derive a complementary relation between the 
excess work and the time lapse.

\section{Brownian Motor}


\begin{figure}[t]
\epsfysize=.18\textwidth
\centerline{\epsffile{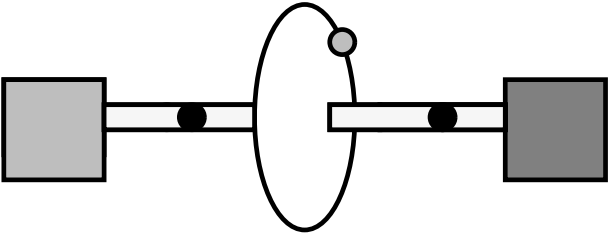}}
\caption{
Schematic figure of a model of Brownian motors driven by particle exchange.
}
\label{fig2}
\end{figure}


\begin{figure}[t]
\epsfysize=.5\textwidth
\centerline{\epsffile{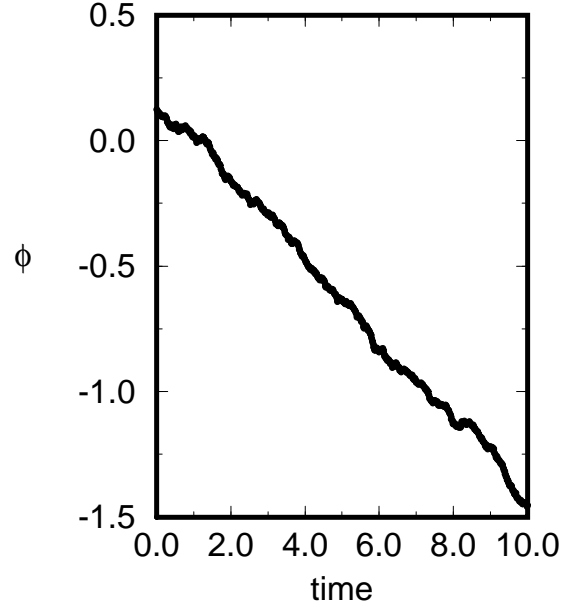}}
\caption{
Averaged time evolution of the motor angle $\phi$.
The graph was created  by averaging $10^4$ samples of $\phi(t)$ with
$t=0.01\times i$, where $1\le i\le 1000$.
}
\label{fig3}
\end{figure}

As an application of the Langevin dynamics in contact with a particle
reservoir, we present a simple model of Brownian motors 
driven by particle exchange. The system is assumed to be
in the region $-L/2 \le x \le L/2$ and to contact with particle reservoirs
at $x=\pm L/2$. The density in the particle reservoirs
are  denoted by $\rho_+$  and $\rho_-$, respectively. 
A motor particle is confined in a circle whose center is located
at $x=0$. (See Fig. \ref{fig2}.) 
We  assume that  time evolution of the angle of  motor angle
$\phi$  is described by a Langevin equation 
with a friction constant $\gamma_\phi$.
The form of the potential function of the system can be chosen rather
arbitrarily. We found however that we need fine tuning of the model so that 
we can confirm the motor behavior numerically.
In this paper, we report the result of the model 
\begin{equation}
U(\{x_i\}, \phi)=\sum_{i=1}^\infty A(x_i)\eta(\phi-\phi_0(x_i)),
\end{equation}
where 
\begin{equation}
A(r)= 
\left\{
\begin{array}{lr}
b &  |r|\le r_0, \\
b ({ (r-r_0)^2 \over (r_1-r_0)^2}-1)^2 &  r_0 \le |r| \le r_1,  \\
0  &|r|\ge r_1, 
\end{array}
\right.
\end{equation}
\begin{equation}
\phi_0(r)= 
\left\{
\begin{array}{lr}
 {\pi \over 2}\left[\left({r\over r_0}\right)^3-3{r\over r_0}\right]
 & |r|\le r_0, \\
 -\pi {\rm sign}(r) &  |r| \ge r_0,  
\end{array}
\right.
\end{equation}
and
\begin{equation}
\eta(\phi)=\sin(\phi)+1.
\end{equation}
Notice that the form of $U(\{x_i\},\phi)$ has been assumed so that
the motor angle turns around when a particle moves slowly from one end
to the other end. 
In numerical simulations, we fix  the values of the following parameters as 
$L=1$, $\gamma=1$, $\gamma_p=0.01$, $b=0.25$, $r_0=0.3$, $r_1=0.4$,
and $\delta t=10^{-4}$, and regard $\rho_+$, $\rho_-$ and $k_B T$ as
control parameters. The chemical potential difference  is then given by
\begin{equation}
\delta \mu=\mu_+-\mu_-=k_B T \log ({\rho_+ \over \rho_-}).
\end{equation}
Under the equilibrium condition $\delta \mu=0$, directed motion
never occur. We confirmed that the averaged time evolution of $\phi$
tends to be constant as the number of samples  increases. 
On the other hand, when $\delta \mu \ne 0$, 
the motor  rotates in one direction on the average. 
In Fig. \ref{fig3}, we showed the averaged time evolution of $\phi$ 
for the parameter values: $(\rho_+,\rho_-)=(2.0,0.04)$ and $k_BT=0.1$.
This shows the clear evidence of  directed motion.  We now discuss
a quantitative aspect a little bit. In particular, we are concerned with 
a ratio of the frequency of the motor rotation  with the particle number 
current at $x=0$, which is denoted by $Y$. Recalling the form of the
potential function,  one may conjecture that 
$Y$ is closed to one.  We found however that $Y$ sharply depends 
on the choice of the parameter value. As one example, in Fig. \ref{fig4},
$Y$ was plotted against $k_BT$. $Y$ decreases quickly for the increment 
of the temperature. We do not understand the reason yet. Detailed study is 
in progress and will be reported elsewhere.


\begin{figure}[t]
\epsfysize=.5\textwidth
\centerline{\epsffile{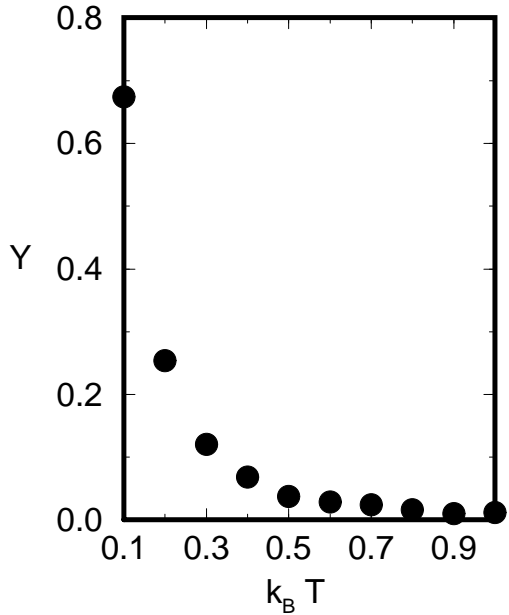}}
\caption{
Ratio of the motor frequency  with  particle number current, denoted by $Y$,
versus $k_B T$.
}
\label{fig4}
\end{figure}

\section{discussion}

We address a few comments. 
Our stochastic model for particle reservoirs seems reasonable, 
but its validity is not  confirmed completely. 
The correspondence between the lattice 
model and the Langevin dynamics  still remains at an intuitive
level.  We expect that a mathematical proof will be presented.
In numerical simulations, we have studied cases 
that there is no interaction among particles. We believe 
that  our model  goes well even if the interaction is included,
because properties of  particle reservoirs should not depend on
the choice of systems. 

We do not understand the nature of Brownian motors so much. 
For example, the work efficiency for Feynmann ratchet was shown
to be much less than Carnot efficiency,\cite{Ken} on the contrary to the
Feynmann's stimulating insight.\cite{Fe}
The efficiency will be discussed in our motor model, but
this will be much less than a value allowed by 
thermodynamics.\cite{shibata} 
Elaborate study on energy transduction along the time axis 
is necessary to clarify the peculiarity of Brownian motor.

The relevance to biological molecular machines may be most
stimulating. Based on the present study, we wish to consider
chemical kinetics,  enzyme catalysis, and  biological membranes.

\acknowledgements
The authors  acknowledge K. Sekimoto, K. Kaneko and Y. Oono 
for their discussions on related topics of nonequilibrium systems.
They  thank T. Chawanya and T. Miyamoto for valuable comments.
They also thank K. Kitahara for sending his unpublished note. 
This work was partly supported by grants from the Ministry of
Education, Science, Sports and Culture of Japan, No. 09740305 
and from National Science Foundation, No. NSF-DMR-93-14938.

\end{document}